\documentstyle[twocolumn,aps,epsfig]{revtex}

\def\sqrtsnn{\sqrt{s_{_{NN}}}}
\def\AuAu{${\rm Au} + {\rm Au}$}

\def\ppbar{${\rm p}\bar{\rm p}$}
\def\avgNp{\langle N_{part} \rangle}
\begin{document}

\title{The significance of the fragmentation region in ultrarelativistic
heavy ion collisions} 
\author{
B.B.Back$^1$,
M.D.Baker$^2$,
D.S.Barton$^2$,
R.R.Betts$^6$,
M.Ballintijn$^4$,
A.A.Bickley$^7$,
R.Bindel$^7$,
A.Budzanowski$^3$,
W.Busza$^4$,
A.Carroll$^2$,
M.P.Decowski$^4$,
E.Garcia$^6$,
N.George$^{1,2}$,
K.Gulbrandsen$^4$,
S.Gushue$^2$,
C.Halliwell$^6$,
J.Hamblen$^8$,
G.A.Heintzelman$^2$,
C.Henderson$^4$,
D.J.Hofman$^6$,
R.S.Hollis$^6$,
R.Ho\l y\'{n}ski$^3$,
B.Holzman$^2$,
A.Iordanova$^6$,
E.Johnson$^8$,
J.L.Kane$^4$,
J.Katzy$^{4,6}$,
N.Khan$^8$,
W.Kucewicz$^6$,
P.Kulinich$^4$,
C.M.Kuo$^5$,
W.T.Lin$^5$,
S.Manly$^8$,
D.McLeod$^6$,
J.Micha\l owski$^3$,
A.C.Mignerey$^7$,
R.Nouicer$^6$,
A.Olszewski$^{2,3}$,
R.Pak$^2$,
I.C.Park$^8$,
H.Pernegger$^4$,
C.Reed$^4$,
L.P.Remsberg$^2$,
M.Reuter$^6$,
C.Roland$^4$,
G.Roland$^4$,
L.Rosenberg$^4$,
J.Sagerer$^6$,
P.Sarin$^4$,
P.Sawicki$^3$,
W.Skulski$^8$,
S.G.Steadman$^4$,
P.Steinberg$^2$,
G.S.F.Stephans$^4$,
M.Stodulski$^3$,
A.Sukhanov$^2$,
J.-L.Tang$^5$,
R.Teng$^8$,
A.Trzupek$^3$,
C.Vale$^4$,
G.J.van~Nieuwenhuizen$^4$,
R.Verdier$^4$,
B.Wadsworth$^4$,
F.L.H.Wolfs$^8$,
B.Wosiek$^3$,
K.Wo\'{z}niak$^3$,
A.H.Wuosmaa$^1$,
B.Wys\l ouch$^4$\\
\vspace{3mm}
\small
$^1$~Physics Division, Argonne National Laboratory, Argonne, IL 60439-4843,
USA\\
$^2$~Chemistry and C-A Departments, Brookhaven National Laboratory, Upton, NY
11973-5000, USA\\
$^3$~Institute of Nuclear Physics, Krak\'{o}w, Poland\\
$^4$~Laboratory for Nuclear Science, Massachusetts Institute of Technology,
Cambridge, MA 02139-4307, USA\\
$^5$~Department of Physics, National Central University, Chung-Li, Taiwan\\
$^6$~Department of Physics, University of Illinois at Chicago, Chicago, IL
60607-7059, USA\\
$^7$~Department of Chemistry, University of Maryland, College Park, MD 20742,
USA\\
$^8$~Department of Physics and Astronomy, University of Rochester, Rochester,
NY 14627, USA\\
}
\date{\today} 
\maketitle

\begin{abstract}\noindent
We present measurements of the pseudorapidity distribution of primary
charged particles produced in \AuAu\ collisions at three energies,
$\sqrtsnn=19.6$, 130, and 200~GeV, for a range of collision
centralities. The centrality dependence is shown to be non-trivial:
the distribution narrows for more central collisions and excess
particles are produced at high pseudorapidity in peripheral
collisions. For a given centrality, however, the distributions are
found to scale with energy according to the ``limiting fragmentation''
hypothesis.  The universal fragmentation region described by this
scaling grows in pseudorapidity with increasing collision energy,
extending well away from the beam rapidity and covering more
than half of the pseudorapidity range over which particles are
produced.  This approach to a universal limiting curve appears to be a
dominant feature of the pseudorapidity distribution and therefore of
the total particle production in these collisions.
\end{abstract}

\medskip

The strong interaction, described by Quantum Chromodynamics~(QCD), may
be studied under conditions of high parton density and high energy
density, using ultrarelativistic heavy ion collisions.  The high
density regime of QCD is sensitive to nonlinear dynamics and
nonperturbative effects, including parton saturation, the onset of
color deconfinement and chiral symmetry restoration. More
specifically, the pseudorapidity density of charged
particles~$dN_{ch}/d\eta$, where $\eta \equiv - \ln \tan (\theta/2)$,
is related to the entropy density at freezeout. It has been
demonstrated that the growth with energy of $dN_{ch}/d\eta$ at
mid-rapidity is modest compared to the original
expectations~\cite{phobosprl,phoprl200}, and provides a strong
constraint on the initial state parton density and further particle
production during the subsequent evolution of the system.  This Letter
focuses on particle production away from mid-rapidity, which
constrains the collision dynamics more completely.

We have measured the pseudorapidity distribution of charged particles,
$dN_{ch}/d\eta$, over a broad range of $\eta$ for \AuAu\ collisions at
a variety of collision centralities (impact parameters). These
measurements were made for three energies, $\sqrtsnn=19.6$, 130, and
200~GeV, covering a span of an order of magnitude in the same
detector, allowing for a reliable systematic study of particle
production with energy in these collisions.  The data were taken using
the PHOBOS apparatus~\cite{phobos2,robert} during the year 2000 and
year 2001 runs of the Relativistic Heavy Ion Collider at Brookhaven
National Laboratory.  The apparatus used in this analysis comprises a
set of silicon detectors with nearly complete coverage for $|\eta| <
5.4$, which are used for detecting the charged particles, and plastic
scintillator counters, covering \mbox{$3<|\eta|<4.5$}, used for
triggering. Event selection cuts were made on the plastic
scintillators or silicon detectors and also on timing information from
the forward hadronic calorimeters, which measure the energy deposited
by spectator neutrons.

The pseudorapidity densities $dN_{ch}/d\eta$ were corrected for
particles which were absorbed or produced in dead material (primarily
the beryllium beampipe or the magnet steel) and for feed-down products
from weak decays of neutral strange particles.  More details of the
analysis procedures leading to the charged particle pseudorapidity
density can be found in Ref.~\cite{phobosdndeta}. Two improvements in
the handling of common mode noise in the silicon detectors have been
implemented for this analysis, leading to small changes in the results
and to a slight reduction in the systematic errors.  First, the
event-by-event common-mode noise in the ring detectors
\mbox{($3.0<|\eta|<5.4$)} was found to grow with pad size, and the
correction scheme was modified to include this effect.  This
refinement was already in place in Ref.~\cite{phoprl200}. Second, for
very high occupancies, the common mode noise correction in the octagon
detector ($|\eta|<3.2$) becomes slightly inaccurate. A comparison of
the data from the octagon detector with those from the more highly
segmented vertex detector was used to determine a correction factor
for this effect as a function of $\eta$, centrality, and beam energy.
This correction was only required near mid-rapidity ($|\eta|<1.5$) for the
central data and it was less than 4\% everywhere even in the 0--6\%
centrality bin of the 200~GeV data. For the 6--15\% bin of the 200~GeV
data and the 0--6\% bin of the 130~GeV data, this correction was less
than 1\% everywhere. For the more peripheral high energy bins and for
the 19.6 GeV data, no correction was required.

The centrality of the collision is characterized by the average number
of nucleon participants~$\avgNp$. For the 130 and 200~GeV data sets,
this was estimated from the data using two sets of 16 paddle counters
covering \mbox{$3<|\eta|<4.5$} forward and backward of the interaction
point. The truncated mean of the signals in each set of detectors is
proportional to the total charged multiplicity in this region of
pseudorapidity. In order to extract $\avgNp$ for a given fraction of
the cross-section, we rely on the fact that the multiplicity in the
paddles increases monotonically with centrality, but we do not assume
that the number of participants is proportional to the
multiplicity. The monotonic relationship between the truncated paddle
mean and $N_{part}$ was verified using the neutral spectator energy
measured in the forward hadronic calorimeters.  For peripheral events,
the dominant uncertainty in $\avgNp$ is given by the trigger
efficiency uncertainty.  Using several methods, based on the HIJING
model~\cite{HIJING}, we have estimated our minimum-bias trigger
efficiency for events with a vertex $z_{vtx}$ near the nominal
interaction point to be \mbox{$\epsilon = 97 \pm 3$\%}.  The final
systematic errors on $\avgNp$, ranging from 3--6\%, are tabulated in
Ref.~\cite{phoboscent200} where this method was outlined in more
detail.

At the lowest RHIC energy, $\sqrtsnn=$~19.6~GeV, the much lower beam
rapidity ($y_{beam} \sim 3$) precludes using the same method of
analyzing the trigger counters (\mbox{$3<|\eta|<4.5$}). Instead, we
construct a different quantity, ``EOCT'', which is approximately
proportional to the multiplicity: the path-length corrected sum of the
energy deposited in the octagon (silicon) detector($|\eta|<3.2$).  In
the first step, both HIJING simulations and simple Glauber
model~\cite{Czyz,Glauber} Monte Carlo simulations are used to estimate
the fraction of the total cross-section ($\epsilon$) in the triggered
sample, as well as to estimate the systematic error. For consistency
with the higher energy results, the low $N_{part}$ tail is assumed to
be distributed as in HIJING, leading to \mbox{$\epsilon=52 \pm 4$\%}.
Once $\epsilon$ is determined, cuts are made in EOCT in a similar way
as with the paddle signal to extract $\avgNp$ for a chosen fraction of
the total cross-section~\cite{phoboscent200}. As with the paddle
signal, we only assume that EOCT is monotonic with $\avgNp$, not that
it is proportional to it.  The $\avgNp$ values and the estimated
systematic uncertainty for various centrality bins in the 19.6~GeV
data are given in Table~\ref{tab:Npart}.

\begin{figure}[htbp]
\centerline{ \epsfig{file=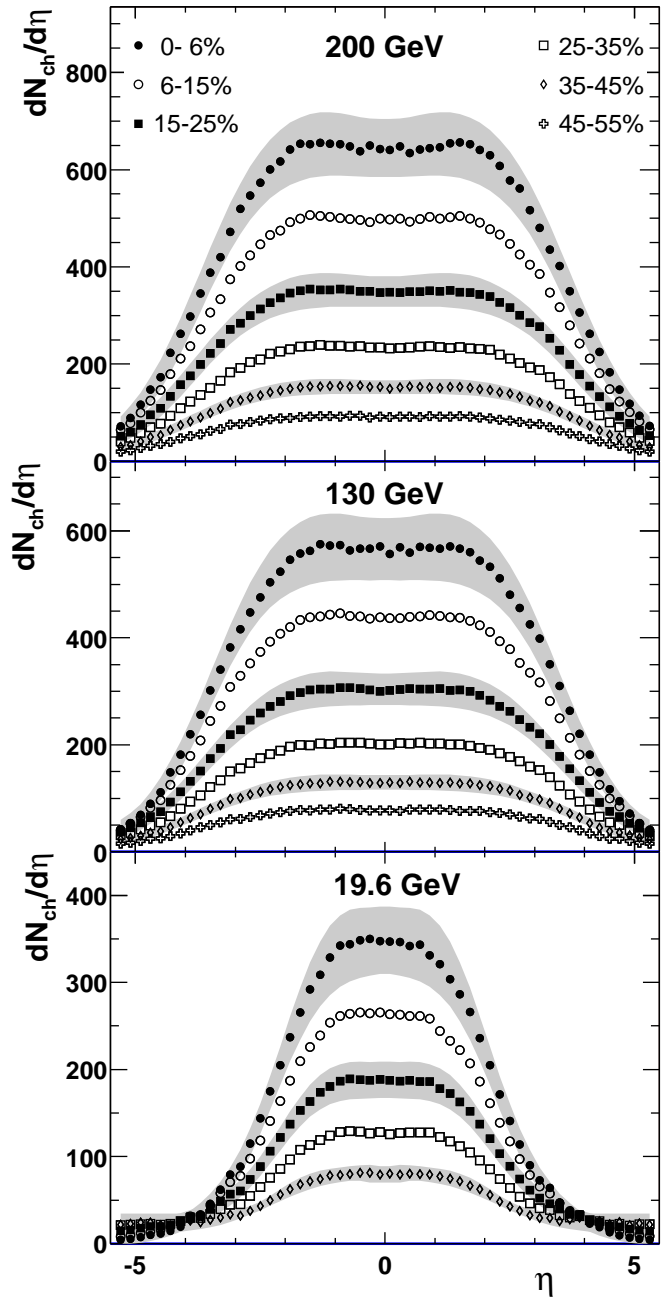,width=8.7cm}}
\caption{The charged particle pseudorapidity distribution,
$dN_{ch}/d\eta$, measured for \AuAu\ at $\sqrt{s_{_{NN}}} =$ 200, 130,
and 19.6~GeV for the specified centrality bins. These bins range from
0--6\% central to 45--55\% in the case of the higher energy data and
0--6\% to 35--45\% for the 19.6~ GeV data.  The statistical errors are
negligible. The typical systematic errors (90\% C.L.) are shown as
bands for selected centrality bins.}
\label{fig:dNdetas}
\end{figure}

Figure~\ref{fig:dNdetas} shows the charged particle pseudorapidity
distributions ($dN_{ch}/d\eta$) measured at $\sqrt{s_{_{NN}}} =$ 200,
130, and 19.6 GeV for different centrality bins for $-5.4<\eta<5.4$.
The statistical errors are comparable to the size of the data points.
The 90\% confidence-level systematic errors are shown as a gray band
for selected, but typical, centrality bins.

Due to the large coverage in $\eta$, $dN_{ch}/d\eta$ is measured over
almost the full range, except for a small missing fraction at very
high $|\eta|$. Using data at lower energy (see Figure~2), this
fraction is estimated to be less than 2\% for the 130 and 200~GeV
central (0--6\%) results.  Table~\ref{tab:Nchtot} lists the integrated
charged multiplicity for three energies in different ranges of
$\eta$. The fiducial range $|\eta|<4.7$ is quoted for easier
comparison to data from the BRAHMS
experiment~\cite{brahmsplb,brahms200}.  After correcting for the
slight differences in centrality binning, the BRAHMS data for 130 and
200~GeV are each $\sim 6$\% below the results quoted here, consistent
within the quoted errors of both experiments. The remaining columns
show the result in the fiducial range $|\eta|<5.4$ of PHOBOS, and the
total inferred $N_{ch}$.

To separate the trivial kinematic broadening of the $dN_{ch}/d\eta$
distribution from the more interesting dynamics, the data for \AuAu\
collisions at different energies can be viewed in the rest frame of
one of the colliding nuclei. Such an approach led to the ansatz of
``limiting fragmentation''~\cite{Yanglimfrag}, which successfully
predicted the energy dependence of particle production away from
mid-rapidity in hadron collisions, including
\ppbar~\cite{UA5limfrag}. This ansatz states that, at high enough
collision energy, both \mbox{$d^2N/dy'dp_{_T}$} and the mix of particle
species (and therefore also $dN/d\eta'$) reach a limiting value
and become independent of energy in a region around $y'\sim 0$, where
\mbox{$y'\equiv y-y_{beam}$} and rapidity \mbox{$y \equiv
\tanh^{-1} \beta_z$}.  



%
Figure~2a shows the scaled, shifted pseudorapidity
distributions \mbox{$dN_{ch}/d\eta'/\langle N_{part}/2\rangle$} for
central \AuAu\ collisions which span a factor of ten in collision
energy ($\sqrtsnn$). The distributions are scaled by $\langle
N_{part}/2 \rangle$ to remove the effect of the different number of
nucleons participating in collisions with different centralities. The
results are folded about mid-rapidity (positive and negative $\eta$
bins are averaged).  The distributions are observed to be independent
of collision energy over a substantial $\eta'$ range.  This is
consistent with and extends a similar observation made by
BRAHMS~\cite{brahms200} over a more restricted $\eta'$ range.  Both
the 19.6 and 130~GeV data reach 85--90\% of their maximum value before
deviating significantly (more than 5\%) from the common limiting
curve.  These data demonstrate that limiting
fragmentation applies well in the \AuAu\ system and that the
``fragmentation region'' is rather broad, covering more than half of
the available range of $\eta'$ over which particles are produced. In
particular, the fragmentation region grows significantly between
19.6~GeV and 130~GeV, extending more than two units away from the beam
rapidity. A similar effect was observed in \ppbar\ collisions, over a
range of $\sqrt{s}$ from 53 to 900~GeV~\cite{UA5limfrag}.  In both
cases, particle production appears to approach a fixed limiting curve
which extends far from the original beam rapidity, indicating that
this universal curve is an important feature of the overall
interaction and not simply a nuclear breakup effect. This result is
in sharp contrast to the boost-invariance scenario~\cite{Bj} which
predicts a fixed fragmentation region and a broad central rapidity
plateau that grows in extent with increasing energy.

\begin{figure}[htbp]
\centerline{ \epsfig{file=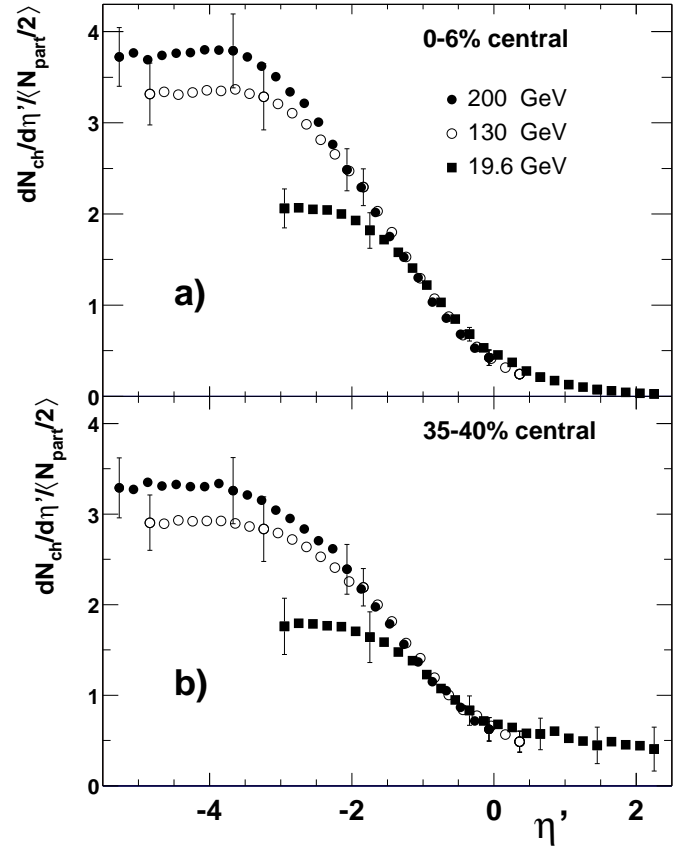,width=9cm}} 
\caption{\AuAu\ data for $\sqrt{s_{_{NN}}} =$ 19.6, 130, and 200~GeV,
plotted as $dN_{ch}/d\eta'$ per participant pair, where $\eta' \equiv
\eta - y_{beam}$ for a) 0--6\% central, b) 35--40\% central.
Systematic errors (90\% C.L.) are shown for selected,
typical, points.}
\label{fig:limfragc}
\end{figure}

Figure~\ref{fig:limfragc}b shows the scaled pseudorapidity
distributions for a set of non-central collisions, which also exhibit
limiting fragmentation over a broad range of $\eta'$.  These results
also show that the extent of the fragmentation region is nearly
independent of centrality.

\begin{figure}[htbp]
\centerline{ \epsfig{file=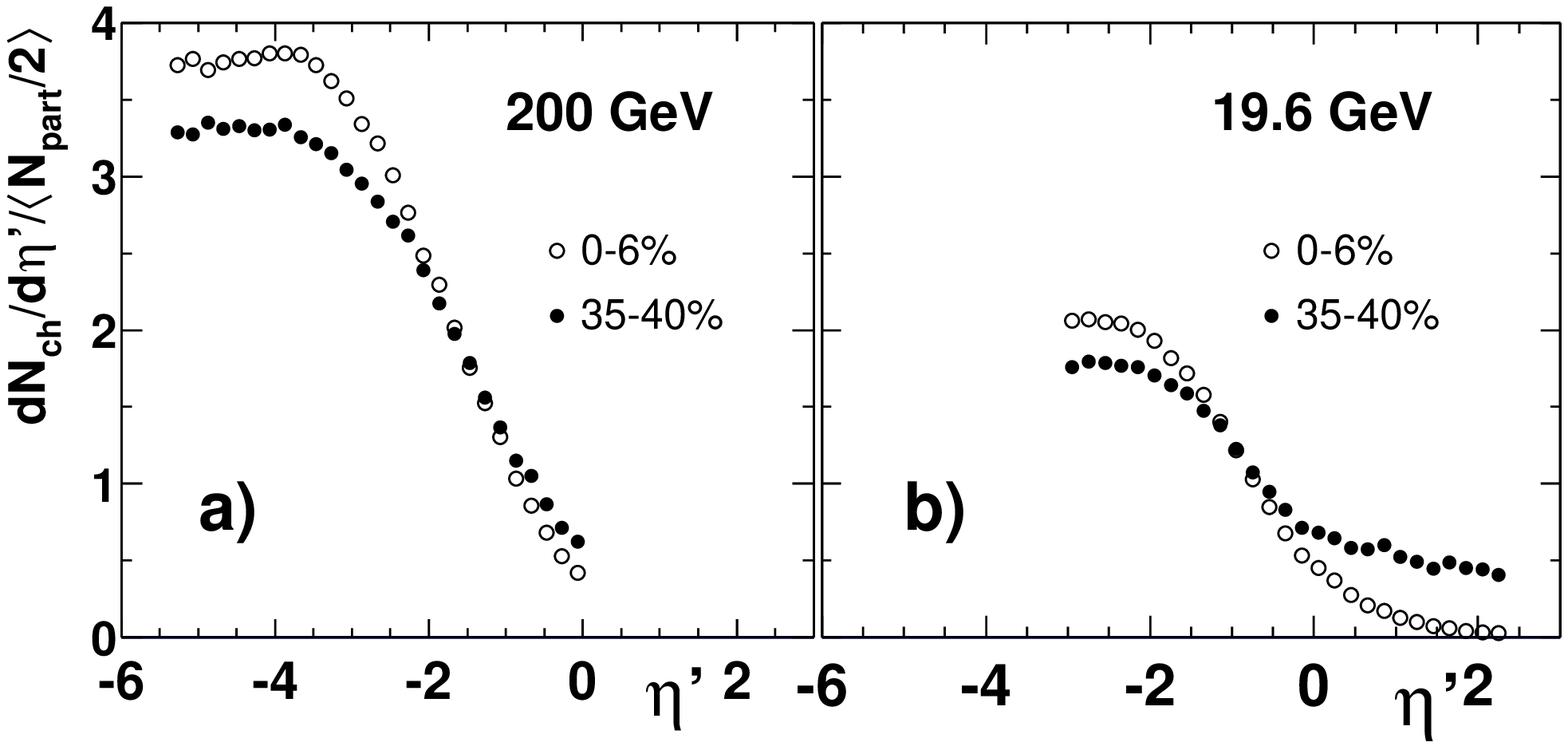,width=9cm}} 
\caption{The distribution $dN_{ch}/d\eta'$ per participant pair for
central (0--6\%) and non-central (35--40\%) \AuAu\ collisions for
a) $\sqrtsnn=200$~GeV, b) $\sqrtsnn=19.6$~GeV.
Systematic errors are not shown.}
\label{fig:centrality}
\end{figure}

\begin{figure}[htbp]
\centerline{ \epsfig{file=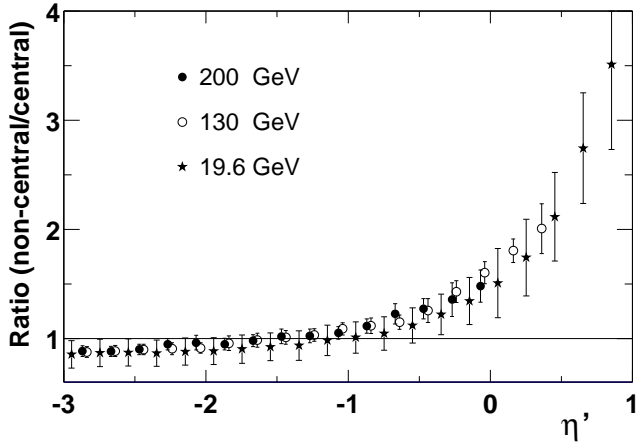,width=9cm}} 
\caption{The ratio of $dN_{ch}/d\eta'$ per participant pair between
non-central (35--40\%) and central (0--6\%) data plotted for $\sqrtsnn=$
200~GeV, 130~GeV, and 19.6~GeV. The errors represent a 90\% C.L.\ systematic 
error on the ratio.}
\label{fig:centratio}
\end{figure}

Figure~\ref{fig:centrality} shows the centrality dependence of the
\mbox{$dN/d\eta'/\langle N_{part}/2 \rangle$} distribution at the
two extreme energies: 19.6 and 200~GeV.  These data demonstrate that
the result in the fragmentation region changes significantly with
centrality.  Figure~\ref{fig:centratio} shows the ratio of non-central
to central data with (90\% C.L.)  systematic errors included. The
error in the ratio involves a partial cancellation of the systematic
errors in the individual measurements.  For $\eta'>-1.5$, the scaled
pseudorapidity density actually grows in the peripheral data with
respect to the more central data.  This effect has already been observed
for \AuAu\ collisions at $\sqrtsnn=$
130~GeV~\cite{phobosdndeta} and for $\rm{Pb}+\rm{Pb}$ collisions at
17~GeV~\cite{EMU13}, and is confirmed here for \AuAu\ at 19.6 and
200~GeV. This contradicts the suggestion, put forward in
references~\cite{brahmsplb,brahms200}, that the limiting curve for
particle production in the fragmentation region is independent of
centrality as well as of energy.  It should be noted that the
hypothesis of limiting fragmentation does not imply that the limiting
curve is independent of centrality, just that $dN/d\eta'/\langle
N_{part}/2 \rangle$ is energy-independent for a fixed centrality, once
at sufficiently high energy.

The strong centrality dependence seen in Figure~\ref{fig:centrality}
has two features: an excess of particles at high $\eta$ in peripheral
\AuAu\ events and a narrowing of the overall pseudorapidity
distribution, perhaps caused by a shift of particles from high $\eta$
toward mid-rapidity for more central collisions. The excess of
particles at high $|\eta|$ can be most easily seen in
Figure~\ref{fig:dNdetas}, where, for the 19.6~GeV data, the absolute
yield of charged particles actually grows for non-central collisions.  It
should be noted that the very peripheral limit of a \AuAu\ collision
is {\em not} an ${\rm N} + {\rm N}$ collision, but rather an ${\rm N}
+ {\rm N}$ collision with two large excited nuclear remnants, one
of which is at $y'=0$. The increased particle production near 
$y \pm y_{beam}=0$ due to the nuclear remnants has been studied in $p+A$
collisions~\cite{pAdata} and for lower energy ${\rm Pb}+{\rm Pb}$
collisions~\cite{EMU13}. The narrowing of the overall distribution may
be due to dynamical effects, such as baryon stopping~\cite{Wit}, or
kinematic effects, such as a shift in $\eta'$ (for fixed $y'$) due to
the particle mix (${\rm p} / \pi$ ratio) changing with
centrality. This narrowing of the distribution can be characterized as
an increase at mid-rapidity~\cite{phoboscent200,phoboscent130} with an
approximately compensating reduction at high $\eta$.

In summary, we have performed a comprehensive examination of the
pseudorapidity distributions of charged particles produced in \AuAu\
collisions at RHIC energies from $\sqrtsnn=19.6$ to 200~GeV, including
an estimate of the full charged particle multiplicity at three
energies. For central collisions at the highest energy, we find that a
total of more than 5000 charged particles are produced.  These results
span eleven units of pseudorapidity, a factor of ten in energy, and a
factor of five in $\avgNp$ --- all measured in a single detector. The
data show a number of interesting features.  First, limiting
fragmentation (energy independence of $dN/d\eta'$) is valid over a
large range of $\eta'$.  Second, the scaled $dN/d\eta$ shape is not
independent of centrality at high $\eta$. The $\eta$ distribution is
broader in peripheral collisions than in central collisions.  Third,
as in \ppbar\ collisions, the fragmentation region in \AuAu\
collisions grows in pseudorapidity extent with beam energy, becoming a
dominant feature of the pseudorapidity distributions at high energy.

\medskip

Acknowledgements: We would like to thank the management of BNL and the
C-A department for providing the variety of collision energies at RHIC
which made this work possible. This work was partially supported by US
DoE grants DE-AC02-98CH10886, DE-FG02-93ER40802, DE-FC02-94ER40818,
DE-FG02-94ER40865, DE-FG02-99ER41099, W-31-109-ENG-38, and NSF grants
9603486, 9722606 and 0072204. The Polish groups were partially
supported by KBN grant 2-P03B-10323. The NCU group was partially
supported by NSC of Taiwan under contract NSC 89-2112-M-008-024.

\begin{table}[htbp]
\caption{Estimated number of nucleon participants in 19.6~GeV \AuAu\ 
collisions according to the centrality bin (quoted as a fraction of the 
inelastic cross-section). The systematic errors are shown. Note: 
the 40--45\% centrality bin has large systematic errors on $\avgNp$,
so $N_{part}$ scaling is not used for that bin. 
\label{tab:Npart}}
\begin{tabular}{rc|rc} 
Centrality & $\avgNp$ & Centrality & $\avgNp$ \\ \hline
 0--3\%:  & $351 \pm 14$  & 20--25\%: & $178 \pm 12$ \\
 3--6\%:  & $322 \pm 10$  & 25--30\%: & $150 \pm 13$ \\
6--10\%:  & $287 \pm 10$  & 30--35\%: & $125 \pm 13$ \\
10--15\%: & $247 \pm 11$  & 35--40\%: & $103 \pm 14$ \\ 
15--20\%: & $210 \pm 12$  &           &              \\ \hline
 0--6\%:  & $337 \pm 12$  & 15--25\%: & $194 \pm 12$ \\
6--15\%:  & $265 \pm 11$  & 25--35\%: & $138 \pm 13$ \\
\end{tabular}
\end{table}

\begin{table}[htbp]
\caption{Total charged multiplicity in three fiducial ranges of $\eta$ for
central (0--6\%) collisions.
\label{tab:Nchtot}}
\begin{tabular}{l|rrr} 
$\sqrtsnn$ & $N_{ch}(|\eta|<4.7)$ & $N_{ch}(|\eta|<5.4)$ 
 & $N_{ch}({\rm total})$ \\ \hline
19.6 GeV & $1670\pm 100$ & $1680\pm 100$ & $1680\pm 100$ \\
130 GeV  & $4020\pm 200$ & $4100\pm 210$ & $4170\pm 210$ \\
200 GeV  & $4810\pm 240$ & $4960\pm 250$ & $5060\pm 250$ \\ 
\end{tabular}
\end{table}


\vspace{-0.6cm}
\begin{thebibliography}{99}
\vspace{-0.5cm}
\bibitem{phobosprl} B.~B.~Back {\it et al.}, Phys.\ Rev.\ Lett.\ {\bf 85}, 
3100 (2000).
\bibitem{phoprl200} B.~B.~Back {\it et al.}, Phys.\ Rev.\ Lett.\ {\bf 88}, 
22302 (2002).
\bibitem{phobos2} R. Nouicer {\it et al.},  Nucl.\ Instrum.\ Methods.\ 
{\bf A461}, 143 (2001).
\bibitem{robert} B.~B.~Back {\it et al.}, Nucl.\ Phys.\ {\bf A698}, 416c 
(2002).
\bibitem{phobosdndeta} B.~B.~Back {\it et al.}, Phys.\ Rev.\ Lett.\ {\bf 87}, 
102303 (2001).
\bibitem{HIJING} M.~Gyulassy and X.~N.~Wang, Phys.\ Rev.\ {\bf D44}, 
3501 (1991).
\bibitem{phoboscent200} B.~B.~Back {\it et al.}, Phys.\ Rev. {\bf C65}, 
061901R (2002).
\bibitem{Czyz} W.~Czyz, L.~C.~Maximon, Annals Phys. {\bf 52}, 59 (1969).
\bibitem{Glauber} R.~J.~Glauber, G.~Matthiae, Nucl.\ Phys.\ {\bf B21}, 
135 (1970).
\bibitem{brahmsplb} I.~G.~Bearden {\it et al.}, Phys.\ Lett.\ B523, 227 (2001).
\bibitem{brahms200} I.~G.~Bearden {\it et al.}, Phys.\ Rev.\ Lett.\ {\bf 88}, 
202301 (2002).
\bibitem{Yanglimfrag} J.~Benecke, T.~T.~Chou, C-N.~Yang, E.~Yen, Phys.\ Rev.\
{\bf 188}, 2159 (1969).
\bibitem{UA5limfrag} G.~J.~Alner {\it et al.}, Z.\ Phys.\ {\bf C33}, 1 (1986).
\bibitem{Bj} J.~D.~Bjorken, Phys.\ Rev.\ {\bf D27}, 140 (1983).
\bibitem{EMU13} P.~Deines-Jones {\it et al.}, Phys.\ Rev.\ {\bf C62}, 
014903 (2000).
(1983) 2580.
\bibitem{pAdata} J.~E.~Elias et al., Phys.\ Rev.\ Lett.\ {\bf 41}, 285 (1978).
\bibitem{Wit} W.~Busza and A.~S.~Goldhaber, Phys.\ Lett.\ {\bf B139}, 
235 (1984).
\bibitem{phoboscent130} B.~B.~Back {\it et al.}, Phys.\ Rev. {\bf C65}, 
031901R (2002).

\end{thebibliography}
\end{document}